\documentclass[preprint,amsmath,amssymb,amsfonts,aps,superscriptaddress,nofootinbib]{revtex4}
\usepackage{ifsym}
\usepackage{MnSymbol}
\usepackage[usenames,dvipsnames]{xcolor}
\usepackage{hyperref}
\def\be{\begin{equation}}
\def\ee{\end{equation}}
\def\ba{\begin{eqnarray}}
\def\ea{\end{eqnarray}}
\def\nl{\nonumber\\}

\def\a{\alpha}

\def\d{\delta}

\def\l{\lambda}

  \def\r{\rho}

\def\cA{{\cal A}}

\def\[{\left[}
\def\]{\right]}
\def\({\left(}
\def\){\right)}

\def\<{\langle}
\def\>{\rangle}

\def\sun{{SU(N)}}  \def\son{SO(N)}

\def\2F1{\,_2{\rm F}_1}

\newcommand{\tr}{\text{Tr}}

\begin{document}

\title{Group-theoretic relations for partial amplitudes in gauge theories with orthogonal group}


\author{Jia-Hui Huang}
\email{huangjh@m.scnu.edu.cn}
\affiliation{Guangdong Provincial Key Laboratory of Quantum Engineering and Quantum Materials,
School of Physics and Telecommunication Engineering,
South China Normal University, Guangzhou 510006,China}


\date{\today}

\begin{abstract}
It is very important to find some nontrivial relations for color-ordered amplitudes at loop levels. In the last several years, a pure group-theoretic method has been proposed to study  loop level relations for color-ordered amplitudes in $\sun$ gauge theories. In this paper, we apply the same method to study the linear relations for loop level four point color-ordered gluon amplitudes in gauge theory with $\son$ group and find that there are $1,6,10,13,16$ constraint relations at $L=0,1,2,3,4$ loop level. It is expected that there are more group-theoretic constraints for high loop levels. The constraint relations should be derived order by order in $\son$ gauge theories. This is unlike the $\sun$ case, where there are only four constraints for higher $(L\geqslant 2)$ loop amplitudes.
\end{abstract}

\maketitle

\section{Introduction}\label{sec:intro}
Scattering amplitudes are important gauge invariant observables in quantum gauge theories. Many effects have been made to simplify the computation of the color-ordered partial amplitudes in recent years. It is found that not all color-ordered partial amplitudes are independent. Tree level partial amplitudes in $\sun$ gauge theory satisfy many relations, such as $U(1)$ decoupling relations\cite{bg1987,mpx1988,bernk1991,FHJ2011,hhj2011}, Kleiss-Kuijf(KK) relations \cite{kk1989,ddm2000} and Bern-Carrasco-Johansson(BCJ) relations\cite{bcj2008,bcj2010,bdv2009,cdf2011}. Due to
the KK relations between color-ordered partial amplitudes, there are $(n-2)!$ independent partial amplitudes for a given helicity configuration and the BCJ relations reduce this number to $(n-3)!$. Just as the tree partial amplitudes, loop level partial amplitudes in $\sun$ gauge theory are also not independent.
One kind of such well-known relations are loop level $U(1)$ decoupling relations, which exist at arbitrary loop level\cite{bernk1991}.  These relations arise from the fact that $U(1)$ photon decouples from pure $\sun$ gluon scattering processes. The KK and BCJ relations also have loop level analogs\cite{ddm2000,bfd2002,cj2011,fjh2012,bi2012, dl2013}. In the past several years, based on a  group-theoretic method\cite{nacu2012,enacu20121,enacu20122}, some nontrivial all-loop relations was proposed for color-ordered loop amplitudes in $\sun$ gauge theories.

Most studies of color-ordered amplitudes focus on $\sun$ gauge theories. One important reason is that $\sun$ gauge symmetry plays important role in modern standard model (SM) of particle physics. It is known that SM suffers from several difficult problems, such as neutrino oscillation and dark matter. Beyond SM, many candidate theories are extensively explored, including grand unified theories (GUT) and string theories. In these theories orthogonal gauge groups often play roles and become important, for example, $SO(10)$ in GUT \cite{am1983,ckn1982,mkfn2002} and $SO(32)$ in string theory\cite{witten1984,nrvw2006,akot2015}. So it is interesting to explore the scattering amplitudes of $\son$ gauge bosons.

In this paper, we consider four point color-ordered partial amplitudes in $\son$ gauge theories. By employing the group-theoretic approach \cite{nacu2012,enacu20121,enacu20122}, we derive the group-theoretic relations satisfied by loop level four point $\son$ partial amplitudes.
 At one loop there are six group-theoretic relations for  $\son$ partial amplitudes. At two and three loop, there are ten and thirteen group-theoretic relations. There are sixteen group-theoretic relations for four loop color-ordered amplitudes. These results are different from the $\sun$ case, due to the different structures of $\son$ and $\sun$ groups. In Section 2, we provide some basic results about $\son$ group and give the trace basis of four point amplitudes. In Section 3, the loop level group-theoretic relations are derived up to four loop level. The last section is the conclusion.

\section{Trace basis of four point amplitudes in $\son$ }
We first list some basic results and our conventions about $\son$ group and algebra. The generators of $\son$ algebra are antisymmetric, traceless Hermitian matrices. In the fundamental representation, the generators can be chosen as follows,
 \ba
 &&l_{ij}=-i(e_{ij}-e_{ji}),i,j=1,2,...,N,\nl
 &&(e_{ij})_{kl}=\d_{ik}\d_{jl}.
 \ea
The above generators can be denoted by $\{T^a\}$ ($a=1,2,...,N(N-1)/2$). They satisfy the $\son$ Lie algebra
 \ba\label{algebra}
 \[T^a, T^b\]=if^{abc}T^c,
 \ea
where $f^{abc}$ are the $\son$ structure constants. They also satisfy normalization condition
 \ba\label{normalization}
 \tr(T^a T^b)=2\d^{ab}.
 \ea
 So, from the above two equations, the structure constant can be expressed as trace of generators
 \ba\label{ftot}
 f^{abc}=-\frac{i}{2}\tr([T^a,T^b]T^c)=-i\tr(T^a T^b T^c)
 \ea
The quadratic Casimir is
 \ba\label{casimir}
 \sum_a T^a T^a=(N-1)I_{N\times N}.
 \ea
Two useful identities for the trace of the generators are
 \ba\label{iden}\nonumber
 &&\tr(T^a A T^a B)=\tr(A)\tr(B)-(-1)^{n_B}(AB^r),\\
 &&\tr(T^a A)\tr(T^a B)=\tr(AB)-(-1)^{n_B}(AB^r),
 \ea
where $A,B$ are series of $\son$ generator matrices or $I_{N\times N}$. $n_{B}$ is the number of generators in $B$ and $n_{B}=0$ if $B=I_{N\times N}$. $B^r$ is product of matrices with the reverse ordering in $B$.

It is known that the scattering amplitudes in $\sun$ gauge theories can be decomposed into two parts: color parts and kinematic parts. The color parts can be expressed in terms of structure constants $f^{abc}$ or traces of $\sun$ matrices in the fundamental representation. This two decompositions can be called $f$-based and trace-based decomposition. In $f$-based decomposition, the $n$-gluon amplitude is written as
 \ba\label{fdecom}
 \cA_n=\sum_\l c_\l a_\l,
 \ea
 where the color basis $\{c_\l\}$ are products of $f^{abc}$ and $\{a_\l\}$ are kinematical factors. In trace-based decomposition, the $n$-gluon amplitude is written as
 \ba\label{tdecom}
 \cA_n=\sum_i t_i A_i,
 \ea
 where the trace basis $\{t_i\}$ are independent traces of $\sun$ matrices and $\{A_i\}$ are the corresponding color-ordered amplitudes. Using some identities of $\sun$ algebra, the color base $c_\l$ can be expanded by $\{t_i\}$,
 \ba
 c_\r=M_{\r i}t_i.
 \ea
 Then, the color-ordered amplitude $A_i$ can be expressed as
 \ba
 A_i=a_\r M_{\r i}.
 \ea
If the matrix $M$ has a nonzero right null vector $r_{k}$, then we can obtain a constraint equation for the color-ordered amplitudes, $A_k r_k=0$.

The above is the basic idea used before to derive group-theoretic constraints for color-ordered amplitudes.  In this paper, we apply the same idea for the four gluon color-ordered amplitudes in $\son$ gauge theories. The trace basis for four gluon amplitudes in $\son$ gauge theories are
 \ba\nonumber
 T_1=\tr(T^{a_1}T^{a_2}T^{a_3}T^{a_4}) ~~~&~~~  T_4=\tr(T^{a_1}T^{a_2})\tr(T^{a_3}T^{a_4}) \\\nonumber
 T_2=\tr(T^{a_1}T^{a_2}T^{a_4}T^{a_3}) ~~~&~~~  T_5=\tr(T^{a_1}T^{a_3})\tr(T^{a_2}T^{a_4}) \\
 T_3=\tr(T^{a_1}T^{a_4}T^{a_2}T^{a_3}) ~~~&~~~  T_6=\tr(T^{a_1}T^{a_4})\tr(T^{a_2}T^{a_3})
 \ea
At tree level, there are only single traces in the trace-based color decomposition. At one loop level, double trace terms appear. For higher-loop amplitudes, trace basis $\{T_i\}(i=1,2,\cdots, 6)$ with different powers of $N$ appear in the decomposition. For four point $L$-loop gauge amplitude, the color decomposition is written as
 \ba\label{decomp}
 A^L=\sum_{m=0}^L\sum_{i=1}^3 N^{L-m}T_i A_i^{(L,m)} +\sum_{n=1}^L\sum_{j=4}^6 N^{L-n}T_j A_j^{(L,n)}.
 \ea
$\{A_i^{(L,m)}\}$ and $\{A_j^{(L,n)}\}$ are the color-ordered amplitudes (partial amplitudes) in $L$ loop with different orders in $N$.

\section{group-theoretic relations for partial amplitudes}
As discussed in the last section, in order to find the group-theoretic relations for partial amplitudes, we should first find out the independent color basis for the four point $L$-loop amplitudes like eq.\eqref{fdecom}, and then find the transformation matrix $M$ between color basis and trace basis. The right null vectors of $M$ give the group-theoretic relations for partial amplitudes.

At tree level, there are three kinds of color basis
 \ba
 c_s=f^{a_1 a_2 b}f^{b a_3 a_4},~~~~~~c_t=f^{a_4 a_1 b}f^{b a_2 a_3},~~~~~~c_u=f^{a_3 a_1 b}f^{b a_2 a_4},
 \ea
which correspond to $s,t,u$-channel diagrams respectively. These color basis are not independent and satisfy Jocobi identity: $c_s=c_t-c_u$. So we can choose $c_s,c_t$ as independent color basis. According to eqs.\eqref{ftot}\eqref{iden}, $c_s,c_t$ can be expanded by $\{T_i\}(i=1,2,3)$. The corresponding transformation matrix is
 \ba\label{zero}
 M^{(0)}=\left(
       \begin{array}{ccc}
         -1 & 1 & 0 \\
         -1 & 0 & 1 \\
       \end{array}
     \right).
 \ea
The right null vector of $M^{(0)}$ is
 \ba\label{r0}
 r^{(0)}=(1,1,1)'.
 \ea
The prime means the transpose of matrix. So the color-ordered amplitudes satisfy
 \ba
 \sum_{i=1}^3 A_i^{(0,0)}=0.
 \ea
This is the dual Ward identity for the tree level $\son$ amplitudes.

For loop amplitudes, it is complicate to explicitly find out the independent color basis. Here we adapt the same assumption used in \cite{nacu2012} that all $(L+1)$-loop
color basis can be obtained from $L$-loop color basis by attaching a rung between two of its external legs. At lower loops, this assumption can be checked explicitly. For a diagram with color factor
 \ba
c=f^{* a_i *}\cdots f^{*a_j*}\cdots,
 \ea
where "$\cdots$" means product of some structure constants, attaching a rung between its external legs $i,j$ means color factor of the resulting diagram becomes to
 \ba\label{rung}
\tilde{c}=f^{b a_i e}f^{e a_j d}f^{* b *}\cdots f^{*d*}\cdots.
 \ea

For four point gauge amplitudes, we assume that $(L+1)$-loop color basis can be obtained by attaching legs $(1,2),(1,3),(1,4)$ from $L$-loop color basis as in eq.\eqref{rung}. Then we consider the effect of this attaching process on the trace basis. Let $ T=\left(
                                                                                                  \begin{array}{c}
                                                                                                    T_1 \\
                                                                                                    T_2\\
                                                                                                    T_3\\
                                                                                                  \end{array}
                                                                                                \right)
$, $\tilde{T}=\left(
                \begin{array}{c}
                  T_4 \\
                  T_5\\
                  T_6\\
                \end{array}
              \right)
$, after the process of attaching a rung, the trace basis transform as follows,
 \ba
 T\rightarrow \(A,B,C\)\left(
                         \begin{array}{c}
                           N T \\
                           T\\
                           \tilde{T} \\
                         \end{array}
                       \right),~~~~~~
 \tilde{T}\rightarrow\(D,E,F\)\left(
                                \begin{array}{c}
                                  N \tilde{T} \\
                                  T\\
                                  \tilde{T} \\
                                \end{array}
                              \right),
 \ea
where $A, B, C, D, E, F$ are all $3\times 3$ matrices and are different for each attaching way (which are given explicitly in Appendix).

Let us explain how to use recursive method to derive higher loop right null vectors from lower loop right null vectors. Suppose $\{c^{(L)}_\a\}$ are the complete (maybe overcomplete) color basis for $L$-loop amplitudes, and they can be expanded by $L$-loop trace basis
$\{T^{(L)}_k,(k=1,2,\cdots,6L+3)\}=\{N^L T_i,N^{L-1} T_i,N^{L-1} T_j,\cdots, T_i, T_j, (i=1,2,3,j=4,5,6)\}$,
 \ba
 c_\a^{(L)}=M_{\a k}^{(L)}T_k^{(L)}.
 \ea
The right null vectors $\{r^{(L)}_k\}$ of $M^{(L)}$ satisfy
 \ba
 M_{\a k}^{(L)}r^{(L)}_k=0,
 \ea
and give the group constraints to $L$-loop color-ordered amplitudes
 \ba
 A^{(L)}_k r^{(L)}_k=0.
 \ea
The $(L+1)$-loop color basis can be obtained by the attaching process and this process also transform the $L$-loop trace basis to $(L+1)$-loop,
 \ba
 T_k^{(L)}\rightarrow G_{kl}^{(L,L+1)}T_l^{(L+1)},
 \ea
where $G$ is $(6L+3)\times(6L+9)$ transformation matrix. The general form of $G$ is
Then we have
 \ba
 c_\a^{(L+1)}=M_{\a l}^{(L+1)}T_l^{(L+1)}= M_{\a k}^{(L)}G_{kl}^{(L,L+1)}T_l^{(L+1)}.
 \ea
The $(L+1)$-loop right null vectors $\{ r^{(L+1)}\}$ satisfy
 \ba
 M_{\a l}^{(L+1)}r^{(L+1)}_l=M_{\a k}^{(L)}G_{kl}^{(L,L+1)}r^{(L+1)}_l=0.
 \ea
The above equation means that $G\cdot r^{(L+1)}$ is a linear combination of $r^{(L)}$. This is the relation between $L$-loop and $(L+1)$-loop right null vectors.

We have obtained $r^{(0)}$ in eq.\eqref{r0}.  The transformation matrix $G^{(0,1)}$ between trace basis of tree amplitudes and one loop amplitudes is
 \ba
 G^{(0,1)}=(A,B,C).
 \ea
By solving recursive equation
 \ba
 G^{(0,1)}\cdot r^{(1)}=\textrm{linear combination of }r^{(0)},
 \ea
we can obtain six one loop right null vectors
 \ba\nonumber
 \{r^{(1)}\}=\{&(1,1,1,0,0,0,0,0,-1)',(0,1,0,0,\frac{1}{2},0,0,0,1)',\\\nonumber
               &(0,0,1,0,0,\frac{1}{2},0,0,1)',(0,0,0,1,1,1,0,0,8)',\\
               &(0,0,0,0,0,0,1,0,-1)',(0,0,0,0,0,0,0,1,-1)' \}.
 \ea
From the last two null vectors, we can see that the one loop color-ordered amplitudes corresponding to double-trace basis are equal to each other. The first four null vectors mean that all one loop sub-leading color-ordered amplitudes can be expressed as linear combinations of leading amplitudes. This is similar with the case in $\sun$ gauge theory. At one loop level, there are nine ($6L+3$) independent trace basis. So, the independent color basis is three.

At two loop, by solving recursive equation
 \ba
 G^{(1,2)}\cdot r^{(2)}=\textrm{linear combination of }r^{(1)},
 \ea
we can get ten two loop right null vectors
 \ba\nonumber
 \{r^{(2)}\}=
             \{&&(10,10,10,1,1,1,0,0,0,0,0,0,0,0,0)',\\\nonumber
               &&(-7,-3,-3,-1,0,0,0,1,0,0,0,0,0,0,0)',\\\nonumber
               &&(-3,-7,-3,0,-1,0,0,0,1,0,0,0,0,0,0)',\\\nonumber
               &&(-3,-3,-3,0,0,0,1,1,1,0,0,0,0,0,0)',\\\nonumber
               &&(40,12,12,8,0,0,0,0,0,1,0,0,0,0,0)',\\\nonumber
               &&(12,40,12,0,8,0,0,0,0,0,1,0,0,0,0)',\\\nonumber
               &&(-16,-16,-16,0,0,0,0,0,0,1,1,1,0,0,0)',\\\nonumber
               &&(26,10,10,4,0,0,0,0,0,0,0,0,0,1,0)',\\\nonumber
               &&(10,26,10,0,4,0,0,0,0,0,0,0,0,0,1)',\\\nonumber
               &&(6,6,6,0,0,0,0,0,0,0,0,0,1,1,1)' \}.
 \ea
These ten null vectors mean ten linear relations between two loop color-ordered amplitudes. It is easy to see that all other color-ordered amplitudes can be expressed as linear combinations of the five amplitudes which correspond to the three leading single-trace basis and two of  the next-to-leading single-trace basis. At two loop, there are 15 independent trace basis and 5 independent color basis.

Using similar procedure, we can obtain the three loop right null vectors
 \ba\nonumber
 \{r^{(3)}\}&&=\\\nonumber
 \{&&(1,0,0,0,0,0,0,0,0,0,\frac{2}{17},\frac{2}{17},0,0,0,\frac{15}{82},\frac{15}{64},\frac{15}{64},\frac{45}{136},\frac{39}{136},\frac{45}{136})',\\\nonumber
 &&(0,1,0,0,0,0,0,0,0,0,\frac{-2}{17},0,0,0,0,\frac{-11}{210},\frac{-8}{77},\frac{-11}{210},\frac{-11}{136},\frac{-11}{136},\frac{-1}{8})',\\\nonumber
 &&(0,0,1,0,0,0,0,0,0,0,0,\frac{-2}{17},0,0,0,\frac{-11}{210},\frac{-11}{210},\frac{-8}{77},\frac{-1}{8},\frac{-11}{136},\frac{-11}{136})',\\\nonumber
 &&(0,0,0,1,0,0,0,0,0,0,\frac{-23}{34},\frac{-23}{34},0,0,0,\frac{-259}{211},\frac{-191}{128},\frac{-191}{128},\frac{-577}{272},\frac{-525}{272},\frac{-577}{272})',\\\nonumber
 &&(0,0,0,0,1,0,0,0,0,0,\frac{23}{34},0,0,0,0,\frac{21}{134},\frac{59}{140},\frac{21}{134},\frac{17}{69},\frac{17}{69},\frac{7}{16})',\\\nonumber
 &&(0,0,0,0,0,1,0,0,0,0,0,\frac{23}{34},0,0,0,\frac{21}{134},\frac{21}{134},\frac{59}{140},\frac{7}{16},\frac{17}{69},\frac{17}{69})',\\\nonumber
 &&(0,0,0,0,0,0,1,0,0,0,0,0,0,0,0,\frac{1}{16},\frac{1}{16},\frac{1}{16},0,\frac{1}{8},\frac{1}{8})',\\\nonumber
 &&(0,0,0,0,0,0,0,1,0,0,0,0,0,0,0,\frac{1}{16},\frac{1}{16},\frac{1}{16},\frac{1}{8},0,\frac{1}{8})',\\\nonumber
 &&(0,0,0,0,0,0,0,0,1,0,0,0,0,0,0,\frac{1}{16},\frac{1}{16},\frac{1}{16},\frac{1}{8},\frac{1}{8},0)',\\\nonumber
 &&(0,0,0,0,0,0,0,0,0,1,1,1,0,0,0,\frac{39}{16},\frac{39}{16},\frac{39}{16},\frac{7}{2},\frac{7}{2},\frac{7}{2})',\\\nonumber
 &&(0,0,0,0,0,0,0,0,0,0,0,0,1,0,0,\frac{-25}{128},\frac{-25}{128},\frac{-25}{128},\frac{5}{16},\frac{-7}{16},\frac{-7}{16})',\\\nonumber
 &&(0,0,0,0,0,0,0,0,0,0,0,0,0,1,0,\frac{-25}{128},\frac{-25}{128},\frac{-25}{128},\frac{-7}{16},\frac{5}{16},\frac{-7}{16})',\\
 &&(0,0,0,0,0,0,0,0,0,0,0,0,0,0,1,\frac{-25}{128},\frac{-25}{128},\frac{-25}{128},\frac{-7}{16},\frac{-7}{16},\frac{5}{16})' \}.
 \ea
 These right null vectors provide 13 constraint equations to the three loop color-ordered amplitudes. This is different from $\sun$ case, where there are only four  constraints at three loop level.  There are 8 linearly independent amplitudes. We can choose 8 most-subleading single-trace and double-trace amplitudes. All other amplitudes can be expressed as linear combinations of them. Alternatively, we can
 also choose 8 leading single-trace and double-trace amplitudes as independent amplitude basis.

At four loop level, we find that the dimension of null right vector space $\{r^{(4)}\}$ is 16. A set of independent null vectors can be chosen as
\ba\nonumber
&(1,0,0,0,0,0,0,0,0,0,0,0,0,0,0,0,\frac{-1}{17},\frac{-1}{17},\frac{-17}{229},\frac{-9}{44},\frac{-17}{229},\frac{-3}{137},\frac{-1}{13},\frac{-30}{389},
\frac{-11}{104},\frac{-8}{33},\frac{-11}{106})'\\\nonumber
&(0,1,0,              0,              0,              0 ,             0,              0,              0,              0,
0,              0 ,             0,              0,              0,              0,              \frac{1}{17},0,\frac{10}{121},\frac{10}{121},
\frac{-1}{21},\frac{1}{599},\frac{ 9}{148},\frac{ 1}{216},\frac{1}{11},\frac{3}{32},\frac{ -8}{165} )'\\\nonumber
&(  0, 0,              1,              0,              0 ,             0,              0 ,             0 ,             0,              0,
 0,              0,   0,              0,              0,              0,              0, \frac{1}{17}, \frac{ -1}{21}, \frac{10}{121},
\frac{ 10}{121},\frac{1}{205},\frac{1}{590},\frac{ 3}{52},\frac{-5}{108},\frac{8}{89},\frac{6}{65})'\\\nonumber
&(0, 0,              0,              1,              0,              0,              0 ,             0,              0,              0,
  0, 0,              0,              0,              0,              0,\frac{31}{68},          \frac{31}{68},\frac{54}{79},\frac{263}{164},
  \frac{54}{79},\frac{35}{166},\frac{34}{55},\frac{139}{224},\frac{27}{28},\frac{153}{79},\frac{88}{93} )'\\\nonumber
&( 0,0,0,   0,              1,              0,              0,              0,              0,              0,
 0, 0, 0,  0,              0,              0,            \frac{-31}{68},0,\frac{-33}{62},\frac{ -33}{62},
 \frac{45}{116},\frac{1}{96},\frac{ -37}{86},\frac{ -1}{96},\frac{ -58}{103},\frac{-69}{118},\frac{ 80}{183})'\\\nonumber
&(0,0,0,0, 0,              1,              0,              0 ,             0 ,             0,
 0, 0, 0,0,              0 ,             0,              0,            \frac{-31}{68},\frac{45}{116},\frac{-33}{62},
 \frac{-33}{62},\frac{-1}{84},\frac{1}{83},\frac{ -15}{37},\frac{11}{26},\frac{-29}{52},\frac{-43}{75} )'\\\nonumber
&(0, 0,0,0,0,              0,              1,              0,              0 ,             0,
 0, 0, 0, 0, 0,              0,              0,              0,            \frac{-31}{357},\frac{ -1}{65},
 \frac{-1}{65},\frac{-1}{135},\frac{-1}{134},\frac{-1}{130},\frac{-5}{87},\frac{-2}{67},\frac{-1}{33} )'\\\nonumber
&(0,0 ,0, 0 , 0,0,              0,              1,              0,              0,
 0, 0, 0,0, 0, 0,              0,              0,             \frac{-1}{65},         \frac{-31}{357},
 \frac{ -1}{65},\frac{-1}{135},\frac{ -1}{134},\frac{ -1}{130},\frac{ -2}{65},\frac{-3}{53},\frac{-1}{33}  )'\\\nonumber
&(0,0, 0, 0, 0, 0,              0 ,             0 ,             1,              0,
 0, 0,0,0,0, 0  ,            0,              0,            \frac{ -1}{65},          \frac{-1}{65},
 \frac{-31}{357},\frac{ -1}{129},\frac{ -1}{140},\frac{-1}{136},\frac{ -2}{77},\frac{-2}{57},\frac{-2}{35})'\\\nonumber
&( 0, 0, 0, 0, 0,              0 ,             0,              0 ,             0  ,            1,
 0, 0, 0, 0 , 0, 0 ,          \frac{ -20}{17},\frac{-20}{17},\frac{-483}{214},\frac{-343}{88},
\frac{ -483}{214},\frac{-65}{82},\frac{ -249}{154},\frac{ -179}{110},\frac{-556}{181},\frac{-495}{103},\frac{-241}{80} )'\\\nonumber
&(0, 0, 0,0,0,0 ,0,  0,              0,              0,
 1,              0 ,             0  ,            0, 0 , 0,\frac{ 20}{17},0, \frac{147}{167},\frac{ 147}{167},
 \frac{ -54}{71},\frac{ -3}{62},\frac{117}{137},\frac{ -1}{66},\frac{417}{469},\frac{68}{73},\frac{-77}{81})'\\\nonumber
&( 0,0, 0,0,  0 , 0, 0,              0,              0,              0,
 0,              1,  0, 0,  0, 0,0,\frac{ 20}{17},\frac{ -54}{71},\frac{ 147}{167},
\frac{ 147}{167},\frac{-1}{90},\frac{-7}{123},\frac{141}{175},\frac{-64}{69},\frac{53}{60},\frac{10}{11} )'\\\nonumber
&( 0,0,0,0, 0,0,              0 ,             0,              0,              0,
 0,0,1, 0,0, 0,              0,              0,            \frac{ 67}{115},          \frac{3}{64},
  \frac{3}{64},\frac{ 1}{42},\frac{ 3}{118},\frac{ 1}{38},\frac{ 33}{119},\frac{27}{257},\frac{5}{47} )'\\\nonumber
&(0,0,0, 0, 0, 0,              0,              0,              0,              0,
 0,0, 0,1,0,0, 0, 0,              \frac{3}{64},          \frac{67}{115},
\frac{  3}{64},\frac{ 1}{42},\frac{ 3}{118},\frac{1}{38},\frac{7}{65},\frac{25}{91},\frac{5}{47}   )'\\\nonumber
&(0,0,0, 0,0,0,0,              0,              0 ,             0,
 0,0,  0, 0,1,   0, 0,              0,              \frac{3}{64},          \frac{ 3}{64},
 \frac{67}{115},\frac{1}{42},\frac{ 2}{77},\frac{2}{77},\frac{ 21}{220},\frac{7}{59},\frac{45}{163}  )'\\
&(0, 0, 0, 0 , 0 , 0,              0,              0,              0,              0,
 0,0,0,0,0, 1,              1,              1,              \frac{8}{3},           \frac{ 8}{3},
\frac{8}{3},          \frac{ 79}{57},\frac{ 151}{114}, \frac{299}{223},\frac{ 194}{57},\frac{166}{51},\frac{10}{3})'
\ea
These sixteen 27-dimensional null vector reduce the number of linearly independent four loop color-ordered amplitudes from 27 to 11. This is different from $\sun$ gauge amplitudes, where there are only four null vectors at four loop level.
The number of linearly independent color basis is 11 at four loop, which is consistent with precious rigorous result\cite{Bern:2010tq}.

\section{conclusion}
In this paper, we consider the group-theoretic constraint relations to four point $\son$ gauge amplitudes up to four loop levels. It is found that at tree level there is one constraint relation for the color-ordered amplitudes and there are $6,10,13,16$ linear constraint relations at $L=1,2,3,4$ loop levels.
 Higher loops group-theoretic relations can be derived similarly. According to our knowledge, these are the first results about color-ordered $\son$ gauge amplitudes. The results are important for particle physics with $\son$ gauge group.

 From the results, we can see that there are similarities and differences between the $\son$ and $\sun$  cases. The number of color basis for them are both $2,3 ,5,8,11$ for $L=0,1,2,3,4$. This is consistent with \cite{Bern:2010tq}. But, the number of trace basis of $\son$ and $\sun$ are respectively $(6L+3)$ and $(3L+3)$ at $L$-loop. The number of $L$-loop color basis of $\sun$ is $(3L-1)(L\geqslant 2)$, so there are four null vectors at all $L(\geqslant 2)$ loop levels. However, for $\son$ case,
 the number of $L$-loop trace basis is $(6L+3)$, so there are more null vectors. This difference obviously comes from the different algebra structures between $\sun$ and $\son$.

{\it \textbf{Acknowledgements:\\}}
The author thanks the anonymous referee for comments and pointing out the paper\cite{Bern:2010tq}.

\appendix*
\section{}
The transformation matrix $G^{(L,L+1)}$ between $L$-loop trace basis and $(L+1)$-loop trace basis  is  $(6L+3)\times(6L+9)$ and has the following form
 \ba
 \left(
   \begin{array}{cccccccc}
     A & B & C & 0 & 0 & 0 & 0 & \ldots \\
     0 & A & 0 & B & C & 0 & 0 & \ldots \\
     0 & 0 & D & E & F & 0 & 0 & \ldots \\
     0 & 0 & 0 & A & 0 & B & C & \ldots \\
     0 & 0 & 0 & 0 & D & E & F & \ldots \\
     \vdots & \vdots & \vdots & \vdots & \vdots & \vdots & \vdots & \ddots \\
   \end{array}
 \right).
 \ea
 For example, the matrix connecting 1-loop and 2-loop trace basis is
\ba
G^{(1,2)}= \left(
   \begin{array}{ccccc}
     A & B & C & 0 & 0 \\
     0 & A & 0 & B & C \\
     0 & 0 & D & E & F \\
   \end{array}
 \right)_{9\times 15}.
 \ea
$A,B,C,D,E,F$ are $3\times 3$ matrices and are of the following forms

 \ba
 A&=&\left(
       \begin{array}{ccc}
         -e_{12}-e_{14} & 0 & 0 \\
         0 & -e_{12}-e_{13} & 0 \\
         0 & 0 & -e_{13}-e_{14} \\
       \end{array}
     \right),\\
  B&=&\left(
       \begin{array}{ccc}
         3e_{12}-2e_{13}+3e_{14} & e_{12}-e_{13} & -e_{13}+e_{14} \\
         e_{12}-e_{14} & 3e_{12}+3e_{13}-2e_{14} & e_{13}-e_{14} \\
         -e_{12}+e_{14} & -e_{12}+e_{13} & -2e_{12}+3e_{13}+3e_{14} \\
       \end{array}
     \right),\\
   C&=&\left(
       \begin{array}{ccc}
         -e_{12}+e_{13} & 0 & e_{13}-e_{14} \\
         -e_{12}+e_{14} & -e_{13}+e_{14} & 0 \\
         0 & e_{12}-e_{13} & e_{12}-e_{14} \\
       \end{array}
     \right),\\
 D&=&\left(
       \begin{array}{ccc}
         -2e_{12} & 0 & 0 \\
         0 & -2e_{13} & 0 \\
         0 & 0 & -2e_{14} \\
       \end{array}
     \right),\\
 E&=&\left(
       \begin{array}{ccc}
         4e_{13}-4e_{14} & -4e_{13}+4e_{14} & 0 \\
         0 & -4e_{12}+4e_{14} & 4e_{12}-4e_{14} \\
         -4e_{12}+4e_{13} & 0 & 4e_{12}-4e_{13} \\
       \end{array}
     \right),\\
 F&=&\left(
       \begin{array}{ccc}
         4e_{12} & 0 & 0 \\
         0 & 4e_{13} & 0 \\
         0 & 0 & 4e_{14}\\
       \end{array}
     \right).
 \ea
In the above matrices, $e_{1i}$ takes one when we attach legs $(1,i)$ and otherwise takes zero.

\end{document}